# gLISA: geosynchronous Laser Interferometer Space Antenna concepts with off-the-shelf satellites

M. Tinto,[1] D. DeBra,[2] S. Buckman,[2] and S. Tilley[3]

[1]*Jet Propulsion Laboratory, MS. 238-737, 4800 Oak Grove Drive, Pasadena, California, 91109, U.S.A.*

[2]*Hansen Experimental Physics Laboratory, Stanford University, Stanford, California, 94305, U.S.A.*

[3]*SSL, 3825 Fabian Way, Palo Alto, California, 94303, U.S.A.*

We discuss two geosynchronous gravitational wave mission concepts, which we generically name gLISA. One relies on the science instrument hosting program onboard geostationary commercial satellites, while the other takes advantage of recent developments in the aerospace industry that result in dramatic satellite and launching vehicle cost reductions for a dedicated geosynchronous mission. To achieve the required level of disturbance free-fall onboard these large and heavy platforms we propose a "two-stage" drag-free system, which incorporates the Modular Gravitational Reference Sensor (MGRS) (developed at Stanford University) and does not rely on the use of µN thrusters.

Although both mission concepts are characterized by different technical and programmatic challenges, individually they could be flown and operated at a cost significantly lower than those of previously envisioned gravitational wave missions. We estimate both mission concepts to cost less than 500M US$ each, and in the year 2015 we will perform at JPL a detailed selecting mission cost analysis.

**I. INTRODUCTION**

The direct observation of gravitational radiation is one of the most pressing challenges in the field of experimental physics of this century. Predicted by Einstein shortly after formulating his relativistic theory of gravity, gravitational waves (GW) will allow us to probe regions of space-time otherwise unobservable in the electromagnetic spectrum[1]. A successful observation will not only represent a great triumph in experimental physics, but will also provide us with a unique test of the various proposed relativistic metric theories of gravity[2,3].

First generation ground-based gravitational wave detectors have been operational for several years. Although they have not been able to detect a gravitational wave signal from several classes of sources searched for, they have identified stringent upper limits to the amplitudes of their expected emitted radiation[4,5,6,7]. It is anticipated that the newly upgraded and more sensitive interferometer detectors will allow us to witness the first detection sometimes in the year 2016 or shortly thereafter. These interferometers operate in the frequency band whose lower limit of about 10 Hz is determined by the sharp rising of the seismic and gravity-gradient noises below this frequency cut-off.

Since the mHz region is expected to be very rich in GW sources characterized by emitting signals of large amplitudes, a natural way to observe them is to build and operate a space-borne detector. The most notable example of a space-based GW interferometer, which has been under study for several decades jointly by scientists in Europe and in the United States, is the Laser Interferometer Space Antenna (LISA) mission. By relying on coherent laser beams exchanged between three dedicated spacecraft forming an almost equilateral triangle of 5 x $10^6$ km arm-length, LISA was designed to detect and study cosmic gravitational waves in the $10^{-4}$ - 1 Hz band[8]. Although over the years only a few space-based detector mission concepts have been considered as alternatives to the LISA mission[9,10,11], with the end in 2011of the ESA/NASA partnership for flying LISA more mission concepts have appeared in the literature[12]. Their goals are to meet most (if not all) the LISA scientific objectives[8] at a lower cost. In this context, two geostationary GW mission concepts[13,14] were independently studied and submitted in response to NASA's Request for Information \# NNH11ZDA019L[15]. Although the scientific capabilities of these missions were shown to compare quite well against those of LISA, the associated cost estimates resulted into figures of about 1B US$, still too high in the current financial environment.

Recent developments in the aerospace industry to meet the growing demands for low-cost satellites and launching vehicles, together with the existing program for flying scientific instruments onboard geostationary commercial satellites[16], have opened up a broader set of opportunities of short development cycle and cost reductions for GW missions. With knowledge of these new opportunities, we have decided to explore the scientific, technical, programmatic, and cost advantages of flying a geosynchronous gravitational wave mission with either "off-the-shelf" satellites or by hosting the necessary instrumentation onboard three geostationary communication satellites, comsats.

The intention of this article is <u>not</u> to present two complete mission studies, but rather to provide an overview of what has already been studied and made available in the open literature[17,18], as well as to discuss some of (what could be regarded as) the "primary" technical road-blocks. The paper is organized as follows. In Section II, after providing a brief description of the hosted payload program onboard commercial geostationary satellites, we address the main technical problem of minimizing the non-gravitational forces acting on these large and heavy satellites. We do this by introducing a new "two-stage" drag-free design, which allows us to achieve the required level of disturbance free-fall needed for detecting GWs with our proposed geostationary mission. Since the gravitational attraction exerted by the comsat on the proof-mass (PM) of our two-stage drag-free system may result into an unacceptably high acceleration noise level, we show that it is possible to uniquely identify the locations and weights of five masses to gravitationally compensate the comsat attraction and reduce the PM acceleration noise to an acceptable level. This scheme could be used by other proposed GW missions relying on spherical proof-masses and the LISA Path Finder mission after a suitable generalization to account for non-spherical proof-masses [19]. Note that for non-spherical PM's, with quadrupole and higher moments, the disturbance forces/torques will have to be computed and either deemed acceptable or compensated.

After noticing that other noise sources and their gradients (electrostatic, electromagnetic, thermal gradient and residual gas, patch effects, electromagnetic interferences from onboard sources, etc.) will need to be carefully studied to quantitatively assess their impacts on the interferometer sensitivity, in Section III we present a geosynchronous mission concept that relies on off-the-shelf components in low cost satellites built by SSL. Besides the obvious and major advantage here of using three dedicated satellites, there exists several other operational advantages that a geosynchronous orbit (with about 5 degree inclination) can offer over the geostationary one such as reduced station-keeping maneuvers and possible cross-links to simplify ground control and mission data systems. In addition, since these platforms are smaller in size and mass than a comsat, a "traditional" drag-free system (i.e. one in which the satellite is continuously centered on a free-floating PM by use of proportional μN thrusters) can be implemented with them. Finally in Section IV we present our final remarks and conclude that each of the mission concepts described in this article should cost less than 500M US$.

**II. A gLISA mission with three geostationary comsats**

The GEOstationary GRAvitational Wave Interferometer mission (GEOGRAWI) was one of two geostationary mission concepts independently submitted in response to the NASA's Request for Information # NNH11ZDA019L[13,14,15] as alternatives to the LISA mission. GEOGRAWI entails three spacecraft in geostationary orbit, forming an equilateral triangle of approximately 73,000 km arm-length. Like LISA, it has three drag-free spacecraft exchanging coherent laser beams but, by being in a geostationary orbit, it achieves its best sensitivity in a frequency band $(3 \times 10^{-2} - 1 \text{ Hz})$[17] that is complementary (in between) to those of LISA and the ground detectors. The astrophysical sources that GEOGRAWI is expected to observe within its operational frequency band include extra-galactic massive and super-massive black-hole coalescing binaries, the resolved galactic binaries and extra-galactic coalescing binary systems containing white dwarfs and neutron stars, a stochastic background of astrophysical and cosmological origin, and possibly more exotic sources such as cosmic strings. GEOGRAWI was deemed capable of testing Einstein's theory of relativity by comparing the waveforms detected against those predicted by alternative relativistic theories of gravity, and also by measuring the number of independent polarizations of the detected gravitational wave signals[2].

To compensate the gravitational perturbations exerted by the Sun, the Moon, and the gravity field of the Earth (which would result in a long-term orbital drift), a geostationary satellite must perform regular operations of "station-keeping"[20]. This entails firing the onboard thrusters to keep the satellite at its required location. We have calculated[18] that the evolution of the GEOGRAWI constellation during the time between two station-keeping maneuvers (about two weeks) is quite benign for the gLISA experiment. In particular, the magnitude of the variations of the inter-spacecraft distances does not exceed 0.05 percent, while the relative velocities between pairs of satellites remain smaller than about 0.7 m/s. In addition it has been calculated that the angles made by the arms of the triangle with the equatorial plane are periodic functions of time whose amplitudes grow linearly with time, and that the maximum variations experienced by these angles as well as by those within the triangle remain smaller than 3 arc-minutes. Finally it has been calculated[18] that the East-West angular variations of the three arms remain smaller than about 15 arc-minutes during a two-week period. It should be emphasized that these relatively small variations of the orbit parameters result in a set of system performance and functional requirements that are less stringent than those characterizing LISA or other heliocentric missions.

To further reduce the costs of a geostationary GW mission we here propose the option of relying on three comsats by taking advantage of the instrument hosting program [16]. Several aerospace companies, such as SSL, Boeing, Lockheed and others, offer to fly for a fee (which can typically vary between 30–70M US$ depending on the instrument's mass, power and complexity) additional payload in the form of scientific instruments on their planned geostationary comsats. They are deployed at a rate of about 20 per year worldwide. Comsat fleet operators will launch to positions around the globe to achieve telecom world coverage. Therefore opportunities exist to host GW payloads on three comsats in three longitudes roughly equally spaced. These comsats have large resources of power and cargo space, especially early in mission life and, as the scientific mission life is less than the 15 year comsat life, it allows the hosted payload to utilize early life system margin. Comsats can weigh more than 5 tons, easily carry 300 kg of additional mass, and transmit the science data to their ground stations at a very high rate. By accessing these platforms we will save the costs associated with the satellites, the launching vehicle, the onboard and ground telecom systems, and minimize part of the engineering costs inherent with the construction and integration of the entire mission.

To take advantage of these satellites for the purpose of detecting and studying gravitational radiation in the mHz band, however, a new drag-free system is needed. A ``traditional'' drag-free design, such as that proposed for the LISA mission, requires precise measurement of the position of the spacecraft relative to a free-floating PM[8,21]. By then acting on the spacecraft with proportional µN thrusters one can maintain the PM in its free-falling state. As coherent laser beams are transmitted/received to/from the other two spacecraft, the PM on each spacecraft provides an almost perfect inertial reference frame with respect to which the frequency fluctuations induced by a GW on the laser light are measured. Although this design has been studied for decades for the LISA mission (and the performance of this subsystem will be experimentally tested on the European LISA Path Finder mission) it can't be made to work onboard a geostationary comsat without proportional thrusters. This is because it would be too difficult and propellant costly to maneuver a comsat around the PM at the required levels of accuracy and precision without proportional thrusters. As it will be shown in the following subsection, we have identified an alternative drag-free design that should allow us to achieve the desired degree of inertia onboard these large, heavy satellites without requiring µN thrusters.

**A. The two-stage drag-free system**

The mass attractions and the other perturbations to the PM are greatest from parts closest to it. Given the large volume available for scientific use onboard a comsat (it can be as large as 1 m x 1 m x 1 m) we have identified an alternative drag-free architecture that provides proportional control to the mass closest to the PM. It can be envisioned by thinking of a small satellite (which from now on we shall refer to as the "hosted satellite"), residing inside an enclosure located on the nadir-pointing (Earth-pointing) deck of the comsat (see Figure 1). It carries onboard the Modular Gravitational Reference Sensor (MGRS)[21] with its free-floating, spinning, spherical PM. Sensors continuously monitor the position and velocity of the hosted satellite relative to the comsat while now, rather than relying on µN thrusters, electro-magnetic actuators attached to the inside walls of the nadir enclosure act on the hosted satellite to keep it centered on its PM (see Figure 2). These actuators can operate within a gap-distance of 1-2 mm. Since the comsat may experience a non-gravitational acceleration of about $10^{-7}$ m/s$^2$ due to solar radiation pressure[22] and disturbances caused by its mechanisms, in order to avoid interruption of data acquisition every 140 s or so as the onboard satellite would move by 1 mm over this time scale, additional thrusters (such as cold-gas thrusters) driven by the onboard sensors can act on the comsat to compensate for the non-gravitational acceleration due to solar radiation pressure. This operational configuration allows the hosted satellite to continuously operate in its drag-free configuration and perform heterodyne measurements by exchanging laser beams with the hosted satellites onboard the other two comsats. This can be done most effectively, though not exclusively, using spherical proof masses[23]. We spin the proof mass to spectrally shift the errors in sphericity and the offset of the geometry center and the mass center of each proof mass. The spherical PM's are 20% hollow to make the principal moment of inertia larger by about 10% than the two other orthogonal and equal moments of inertia so that the polhode frequency is as high as 0.1 of the spin speed and outside the measurement frequency range. In this way we avoid the complexity of the control of the extra degrees of freedom that are required if a non-spherical PM is used.

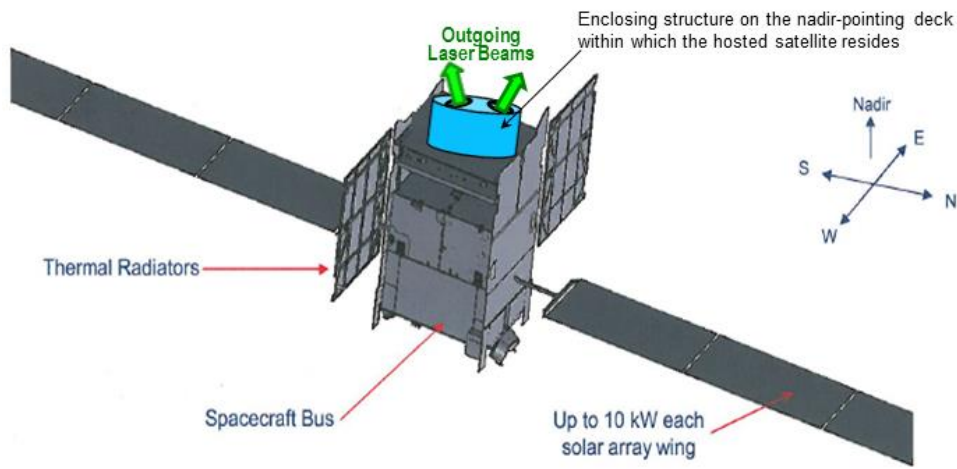

FIG. 1. Example of a commercial geostationary satellite to be used by gLISA. The nadir-pointing deck provides the necessary space for an additional enclosure added to it (blue-colored, online version) where the "hosted satellite" with its gravitational reference sensor will be housed. Laser light beams (green arrows, online version) are exchanged with the other two hosted satellites.

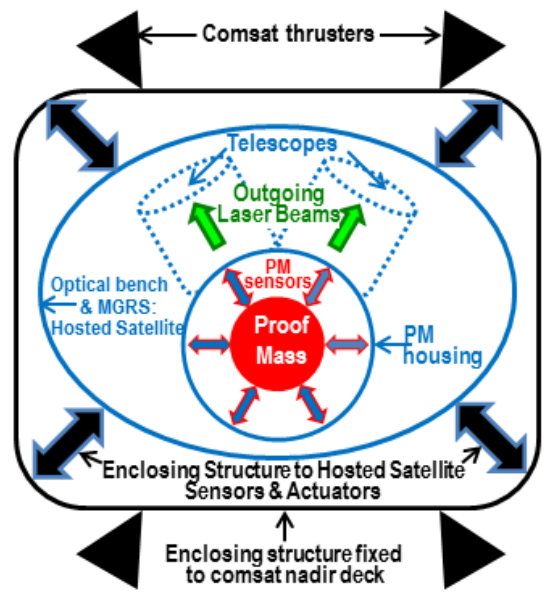

FIG. 2. The "hosted satellite", containing the optical bench and the Modular Gravitational Reference Sensor (MGRS), is freely-floating inside the enclosing structure onboard the comsat. Sensors and actuators (attached to the enclosing structure) are used to measure the position of the hosted satellite and keep it centered on the spherical proof-mass respectively. Since the gap between the actuators and the hosted satellite can be of a few mm, one can maintain such a distance by acting on the comsat with cold-gas (instead of μN) thrusters to compensate for the effects of non-gravitational forces (primarily solar radiation pressure).

The drag-free design sketched above in Figure 2, i.e. the "two-stage drag-free system", is the core part of the instrumental configuration we envision for the geostationary GW mission that relies on comsats. The first stage is the slow 6 degrees of freedom (DoF) control loop that maintains the comsat to hosted satellite position. Its dead band is about 1 mm and its actuators are cold gas thrusters hosted on the comsat. There is no need for μN trusters, which have been too slow to reach maturity for this type of application. The first stage sensors for 5 DoF (3 translations and two rotations), schematically shown in Figure 2, measure the comsat position with respect to the hosted satellite, while the 6$^{th}$ DoF, one rotation, is determined by the orientation of the comsat's nadir platform. The second stage uses fast proportional actuators, for example electromagnetic actuators like those used in LIGO, to maintain the position of the PM housing (attached to the hosted satellite) with respect to the PM to a precision of 1 – 10 nm (see Figure 2). The second stage 3 translational DoF sensing system is the Differential Optical Shadow Sensor System (DOSS), part of the MGRS, with the 3 rotational DoF provided by the two incoming laser beams from the other satellites in the triangular formation. Note that, as the two laser beams require 4 and not 3 DoF, the additional 7$^{th}$ DoF is realized by adjusting the angle between the two telescopes.

By using electromagnetic actuators for the proportional high-accuracy control of the inner-part of the hosted satellite (where the mass attraction is most critical), we avoid the need for μN thrusters, which is a less proven technology for this application. Several noise sources will ultimately define the acceleration noise level of our proposed two-stage drag-free system. Sources such as electrostatic and electromagnetic effects, thermal gradient and residual gas, patch effects, electromagnetic interferences from onboard sources, etc., could potentially degrade the sensitivity of our two-stage drag-free system to unacceptable levels if no adequate electromagnetic and thermal shielding are implemented. Among all the noise sources, however, the comsat-induced gravitational acceleration noise turns out to be the most serious one requiring special attention. In the following subsection we will discuss in detail this noise source and propose a way to suppress its magnitude to a level compatible to the scientific objectives of our geostationary GW mission.

**B. Comsat-induced gravitational noise and its compensation**

Based on a model of the mass distribution of the SSL 1300 satellite platform, in which each spacecraft component is approximated as having constant mass density, we have numerically estimated the magnitudes of the gravitational acceleration and gravity-gradient experienced by the PM of our two-stage drag-free system added on the comsat nadir-pointing deck. We found them to be unacceptably large, with magnitudes of about $1.4 \times 10^{-8}$ m/s$^2$ and $8.7 \times 10^{-9}$ s$^{-2}$ respectively. In order to reduce these values to levels compatible with the acceleration noise requirements of our GW mission (i.e. the square-root of the acceleration noise spectrum to be equal to $3.0 \times 10^{-15}$ m/s$^2$ (Hz)$^{-1/2}$ in the observational frequency band (~ $10^{-2}$ – 1) Hz) additional masses need to be added onboard to compensate for the "gravitational attraction" of the comsat. These masses have to be specified in number, and each of them has to be characterized in magnitude, shape, and location. For the purpose of answering these questions and test quantitatively the validity of the solutions through numerical analysis, we have assumed perfect knowledge of the comsat mass distribution. This allows us to estimate the exact values of both the gravitational acceleration, $\vec{a}$, and gravity gradient, $\overline{\overline{gg}}$ exerted by the comsat at the PM nominal location, $\vec{o}$.

In a coordinate system of choice associated with the comsat, $[x, y, z]$, the location of point $\vec{o}$ has been specified by the following position vector:

$$\vec{o} = [0, \ 1.27, \ 3.81] \text{ m} \tag{1}$$

while, at point $\vec{o}$, the gravitational acceleration, $\vec{a}$, and gravity gradient, $\overline{\overline{gg}}$, have the following components (again in the comsat coordinate system)

$$\vec{a} = [9.12 \times 10^{-8} \quad -0.77 \quad -1.15] \times 10^{-8} \text{ m·s}^{-2} \tag{2}$$

$$\overline{\overline{gg}} = \begin{bmatrix} -4.77 & -5.35 \times 10^{-7} & -6.10 \times 10^{-7} \\ -5.35 \times 10^{-7} & 1.84 & 6.31 \\ -6.10 \times 10^{-7} & 6.31 & 2.93 \end{bmatrix} \times 10^{-9} \text{s}^{-2} \tag{3}$$

In the above numerical expressions we may notice that the x-component of the acceleration and the (xy, xz) components of the gravity gradient are more than seven orders of magnitude smaller than their remaining components. This is because point **o** is located in the comsat (zy) mass distribution symmetry plane.

To identify the number of compensating masses to be added onboard the comsat, we should first remind ourselves that the gravitational gradient, $(gg)_{ij}$, is equal to the second partial derivatives of the gravitational potential, $V(r)$. $(gg)_{ij}$ is therefore a symmetric tensor whose trace is equal to zero as a consequence of the Poisson's equation in vacuum, $\nabla^2 V = 0$. This implies that, in the coordinate system in which the gravity gradient is diagonal, say (X, Y, Z) (i.e. in the coordinate system determined by the eigenvectors of the gravity gradient), only two of its three eigenvalues are independent since their sum must be still equal to zero. Since the acceleration still has 3 independent components in this new coordinate system, and the gravity gradient has only 2, we may argue that to simultaneously cancel the acceleration vector and the gravity gradient at point **o** we need 5 additional masses. Before proceeding with our problem, it may be helpful to write down explicitly the numerical values of the components of the acceleration and gravity gradient in this new (X, Y, Z) orthonormal coordinate system

$$\vec{a'} = [-4.29 \times 10^{-9} \quad 0.21 \quad -1.37] \times 10^{-8} \text{ m·s}^{-2} \tag{4}$$

$$\overline{\overline{gg'}} = \begin{bmatrix} -4.77 & 0 & 0 \\ 0 & -3.95 & 0 \\ 0 & 0 & 8.72 \end{bmatrix} \times 10^{-9} \text{ s}^{-2} \tag{5}$$

In order to identify the location and values of the compensating masses, for simplicity we have assumed them to be of spherical shape and constant mass density. In the coordinate system given by the eigenvectors of the gravity gradient, it is then easy to see that the directions along which the 5 masses will have to lie on coincide with the three coordinate axes (X, Y, Z). In particular, if we add two pairs of spheres on the X− and Y− axes respectively and located in such a way to "bracket" point *o*, we should be able to compensate both the acceleration and gravity gradient components along these two directions. The remaining mass will instead need to be located on the positive Z− axis in order to counter-balance the negative Z− component of the gravitational acceleration.

By fixing the distance to point *o* of each compensating mass, we can then solve for the values of the masses (which allow simultaneous cancellation of the acceleration and gravity gradient at point **o**) by solving a non-homogeneous linear system of 5 equations in 5 unknowns. Note that the closer the compensating masses can be to point *o*, the lighter their values will result, but then the more sensitive the disturbance will be to the relative motion. In what follows we briefly summarize the results of one particular configuration, in which all compensating masses are at a distance of about 25 cm from point **o**. Although in the final onboard flying configuration it might happen that the mass distances from point *o* will be different from this value, our choice allows us to get some insight on the magnitude of the compensating masses.

We found that the two compensating masses located on the X− axis to be essentially equal to each other and to a value of 7.1 kg. This is because the X− component of the acceleration is approximately zero. The masses on the Y− axis turned out to be appreciably different from each other as they had to account for larger acceleration and gravity gradient along this direction. Their values have been estimated to be equal to 6.1 and 8.2 kg. Finally, the mass along the Z− axis resulted to be equal to 13.2 kg, a larger value than the others since it needs to cancel a larger component of the acceleration along the −Z direction. These mass values indicate that a modest amount of onboard additional mass will allow us to simultaneously cancel the comsat acceleration and gravity gradient at the PM nominal location. As noted above, additional compensation might be needed for a non-spherical PM.

Although our procedure results in a perfect cancellation of the comsat-induced gravitational acceleration and gravity-gradient at the nominal PM location, finite accuracy in the knowledge of the comsat mass-distribution together with the 1 mm "dead-band" within which the comsat will be moving relative to the PM, may still result in an above specs residual PM acceleration. In this eventuality, to further reduce this remaining acceleration below the level compatible with the mission science objectives, we may take advantage of the sensors exquisitely accurate and precise measurements of location and velocity of the hosted satellite (and therefore of the PM) relative to the comsat. These additional data streams allow us to reconstruct the effects of the residual PM acceleration imprinted in the TDI measurements[24] and suppress them to the required level.

**III. A gLISA mission with off-the-shelf satellites**

We anticipate proposing two types of mission concepts for 3 satellites: geostationary and other orbit patterns. One challenging point of the previous GW mission concept that relies on hosting on geostationary comsats is that three GW payloads must be spread about the GEO arc and then maintain coordinated operations during science observations. The equilateral triangle need not be strictly followed, but no laser path can approach grazing the Earth limb and rough alignments of laser lines of sight will be fixed before launch. The timing of three launches to three slots may span several years, so some GW payloads might be kept in onboard storage-mode initially. Given the required simultaneous quiet environment for GW observations, spacecraft thruster operations for momentum management and station keeping must be coordinated on all three spacecraft. The most likely scenario is working with a global satellite fleet operator to procure, launch and operate three satellites in this manner. For these reasons there are of course advantages in designing and operating a dedicated GW mission (i.e. one relying on dedicated satellites) if the cost challenges can be mitigated.

Geostationary orbits have a 24 hour period and nearly zero inclination, eccentricity and east-west drift to allow them to "hover" directly above the same equatorial longitude and appear fixed in the sky to a ground user, which simplifies ground terminal designs. Geosynchronous orbits allow modest daily motion to occur as long as it repeats over 24 hours. A dedicated GW mission concept can allow inclination to vary as this offers advantages to the spacecraft design to minimize complexity and cost.

Our dedicated mission concept uses one launcher to insert a main spacecraft carrying two free-flyers to a Geosynchronous Transfer Orbit (GTO). The main spacecraft includes a liquid propulsion subsystem to deliver it to an inclination drift orbit. Sizing of the main satellite and free flyers are driven by the main requirements for the GW payload given in Table I. Other more detailed requirements like magnetic cleanliness and contamination control will need to be developed as instrument hosting requirements. The main spacecraft has the typical subsystems as a standard bus in the SSL 1300 class with 5 kW of power, while the free-flyers are smaller spacecraft with simplified propulsion and power subsystems. The free-flyers are similar to the bus developed by SSL for the Globalstar constellation in the 300 kg mass and 1.5 kW power range. The preliminary mass allocation of the entire separated system is shown in Table II. Chemical propellant (NTO/MMH) budget parameters assumed a SpaceX Falcon 9[25] launch to GTO, main spacecraft orbit rising to geosynchronous inclination drifter orbit and 5 year mission. Gaseous nitrogen (GN2) budgets included attitude control and solar pressure corrections for a 5 year mission.

Table I. Preliminary Spacecraft Hosting Requirements for GW Payload

| Mass | 200 kg |
|---|---|
| Power | Less than 500 W avg. operating, and less than 300 W survival. |
| Size | 1m x 1m x 0.5m volume |
| Thermal Baseplate | $10^0$ C to $30^0$ C operating, $-10^0$ C to $45^0$ C non-operating |
| Mission data rates | 1 Mbps avg., 10 Mbps peak |
| Housekeeping T&C | 10 kbps telemetry, 1 kbps command |

Table II. Spacecraft Separated Mass Breakdowns (Total of 4620 kg)

| Spacecraft | Bus (kg) | Payload (kg) | Propellants (kg) |
|---|---|---|---|
| Main Spacecraft | 1200 | 200 | 2100 (NTO/MMH), 60 (GN2) |
| Free Flyer 1 | 300 | 200 | 30 (GN2) |
| Free Flyer 2 | 300 | 200 | 30 (GN2) |
| Totals | 1800 | 600 | 2220 |

The following paragraphs illustrate the launch and deployment of a GW geosynchronous mission. The single launch is a key cost limiting item, hence the main spacecraft plus free-flyers is baselined. A Falcon 9 launch has been assumed given its reduced launch cost compared to current Evolved Expendable Launch Vehicle program (EELV) or Ariane launch costs. Proper selection of mission inclination reduces system cost by minimizing the on-orbit spacecraft propulsion requirements. A technique known as "inclination drifting" is used to allow spacecraft to maintain an inclination range without use of any significant propulsion system. For our GW mission concept, selection of a Right Ascension of the Ascending Node (RAAN) in the 253-330 deg range allows inclination to start at and stay below 3 deg for at least 5 years, as illustrated in Figure 3.

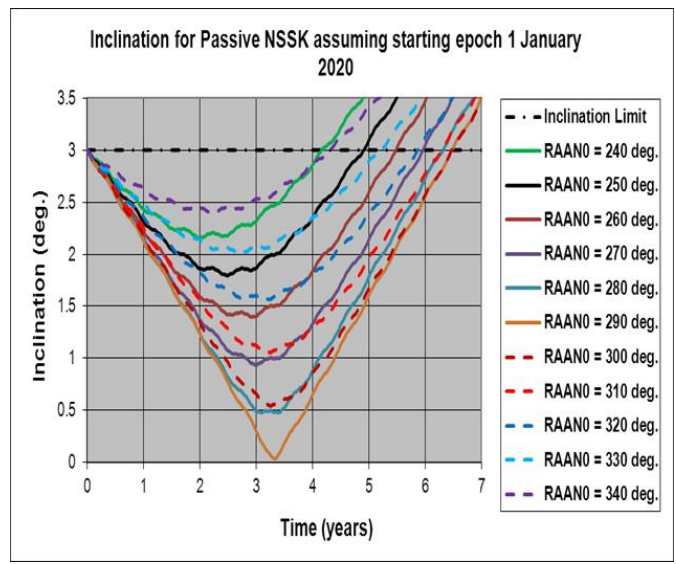

FIG. 3. RAAN Selection and Inclination Drift for 3 Years

After launch vehicle separation, the main spacecraft carries two free-flyers during the orbit circularization phase and then drifting around the orbit to release them roughly 120 deg apart before drifting another 120 deg to its slot within sight of the primary mission ground station. This process takes approximate 2 months. At the target inclination and RAAN, all three spacecraft will be operated in the same orbit plane. Once on-orbit, all three spacecraft fly in relative formation with minimal propulsion activity. The on-board nitrogen propulsion system will be active on all three satellites to counteract solar radiation pressure. At the end of the mission, a small amount of propellant will be used to raise the spacecraft 300 km above the GEO altitude to a graveyard orbit.

Since ground operations can be a significant portion of total mission costs, the dedicated system uses cross-links to allow one central gateway and mission control center that coordinates all activities. Heritage Ka-band systems with 0.75 m reflectors exist, providing an up to 10 Mbps peak data rate. The free-flyers maintain a minimal omni Telemetry & Command (T&C) capability for raising the orbit and contingencies, but all routine operations have T&C and mission data routed through fixed crosslinks to the main spacecraft for relay to the central ground control center. Orbit determination may be done by infrequent ranging, on-board GPS or crosslink microwave ranging. Time coordination is controlled by the main spacecraft.

Science data collection is controlled through ground generated instrument operations plans that are uplinked to the spacecraft weekly and stored on-board for time tagged execution. Science data return is maximized with > 90% collection times on a daily basis except for eclipse seasons at the equinoxes. To avoid costly design features for thermal accommodations for science during eclipses, the payloads will be shut down for 45 days each spring and fall. This will simplify power requirements for survival rather than science operations during eclipses. It also avoids time when the sun will interfere with each of the three laser legs. Hence a 4 year on-orbit period will yield 3 years of science data. During 9 months of each year, science will be maximized by coordinating all activities on the three satellites to have similar maneuver times and similar housekeeping periods to allow 2 to 20 hour measurement windows. In theory, GW detection windows could extend to multiple days. Routine scheduling of activities could allow a 30 minute unload window every other day and a 2 hour east-west orbit trim every 2 weeks. Large orbit formation adjustments could be done during the spring or fall eclipse seasons while the science payload is off. Caging of the PM would be required during any significant orbit adjust and during the eclipse season when the solar radiation pressure compensation is off.

Mission cost is a significant driver of this concept. This dedicated mission allows certain aspects to be tailored to our GW experiment without concern for a satellite operator with a main commercial payload that must be continually operating during the GW mission and beyond. The main cost reduction features this dedicated system can leverage are summarized below:

- Single launch, notionally on Falcon 9
- Choice of inclination drift mission orbit to improve dry mass delivered to orbit and reduce propulsion subsystem costs and on-orbit operations
- Main spacecraft to provide orbit positioning for two free-flyers to simplify free-flyers
- Simplified low mass, low power free-flyers with minimal propulsion to host 2 of 3 GW payloads
- Main spacecraft to provide data relay with two free-flyers to simplify ground control operations

- Use of single ground control station to coordinate activities on three spacecraft and collect all mission data
- Eclipse season science data collection stand down to avoid complicated power and thermal solutions to maintain payload operations during challenging thermal environments and avoid sun interferences on laser line of sight

These spacecraft and launch costs are largely known within the commercial satellite industry as the vast majority of bus components are off the shelf with flight heritage. Each of these items listed above will significantly contribute to reducing the mission cost of a dedicated system. During development of this type of programs, other cost saving opportunities will likely arise. Large cost growth in the areas of spacecraft, launch and ground control is unlikely given the maturity of these types of systems.

Further mission cost offset potential is associated with using the main spacecraft as a host for other revenue generating payloads. After the GW mission is over in 5 years, this satellite has tremendous residual capabilities to continue hosting other science payloads for another 5-10 years with pre-planned added propellant loading. The satellite can provide power, thermal, and pointing as required along with a mission data link to the ground. It could also serve as a comsat with a modest (predetermined or reconfigurable) RF payload that could be re-positioned to an orbital slot with commercial service demand. These notional post-GW mission uses indicate the potential for unlocking the residual value of the main satellite to further reduce the GW mission program cost.

**IV. Conclusions**

We have presented two geosynchronous GW missions concepts that rely on off-the-shelf spacecraft technologies. One takes advantage of the science instrument hosting program on geostationary commercial satellites while the other relies on three dedicated, low-cost, satellites. To achieve the required level of disturbance free-fall onboard these large and heavy platforms we have identified a "two-stage" drag-free system, which incorporates the Modular Gravitational Reference Sensor (MGRS) (developed at Stanford University) and does not rely on the use of μN thrusters.

We expect both mission concepts to be significantly cheaper than previously proposed GW missions. This is because, in the case of adding GW payloads onboard three commercial comsats, we will leverage on the entire commercial infrastructure by paying a relatively small "flying" fee. The dedicated-satellite mission concept instead will take advantage of new satellite technologies developed for the commercial aerospace industry together with a relatively new launching vehicle capability (Falcon 9) whose cost is significantly smaller than "traditional launchers" because of its recycling program[25]. Although we have estimated both mission concepts to cost less than 500M US$ each, in the year 2015 we will perform at JPL a detailed TEAM-X[26] selecting mission cost analysis.

**ACKNOWLEDGMENTS**

Massimo Tinto thanks Dr. Anthony Freeman for stimulating conversations on the instrument hosting program onboard commercial geostationary satellites, and acknowledges financial support provided by the Jet Propulsion Laboratory Research & Technology Development program. Daniel DeBra and Sasha Buchman appreciate the support from the KACST-Stanford Center for Excellence in Aeronautics and Astronautics. Scott Tilley acknowledges fruitful discussions with Alfred Tadros, Peter Lord, and Andy Turner at SSL. For Massimo Tinto, this research was performed at the Jet Propulsion Laboratory, California Institute of Technology, under contract with the National Aeronautics and Space Administration.